\documentclass{PoS}

\title{Heavy Elements in the Early Galaxy}

\ShortTitle{Heavy elements in the early Galaxy}

\author{\speaker{Terese Hansen}%
\\
        University of Heidelberg, ZAH, LSW\\
        E-mail: \email{thansen@lsw.uni-heidelberg.de}}

\author{Johannes Andersen\\
        Niels Bohr Institute, University of Copenhagen\\
        E-mail: \email{ja@astro.ku.dk}}

\author{Birgitta Nordtr\"{o}m\\
        Niels Bohr Institute, University of Copenhagen\\
        E-mail: \email{birgitta@nbi.dk}}

\abstract{The oldest stars in the universe retain to a great extent detailed
  information on the chemical composition of the interstellar medium at the
  time of their birth. Hence the earliest phases of Galactic chemical
  evolution and nucleosynthesis in the early universe can be investigated by
  means of the old metal-poor stars. For the majority of extremely metal-poor
  (EMP) stars the element ratios follow a normal pattern, but 10-15\% of the
  stars are enhanced in heavy neutron-capture (\emph{r}- or
  \emph{s}-processes) elements by large factors, and about 20\% are strongly
  enriched in carbon. The enhancement of some elements could be the result of
  highly non-spherical supernova and inefficient mixing in the early
  interstellar medium. Alternative, they could be due to mass transfer from a
  former AGB or supernova binary companion that has now evolved to a white
  dwarf or neutron star. If the latter explanation is true we can detect these
  stars as long-period binaries. Radial velocity monitoring over a period of
  $\sim$5 years of a sample of EMP stars with either \emph{r}-process element
  and/or carbon enhancement is presented. The results indicate that pure
  \emph{r}-process and carbon enhancements are not results of mass transfer
  from a binary companion.}

\FullConference{XII International Symposium on Nuclei in the Cosmos\\
                 August 5-12, 2012\\
                 Cairns, Australia}

\begin{document}

\section{Introduction}
Extremely metal-poor stars should represent the composition of matter in
early galaxies. Thus, from detailed abundance analysis of such stars, we can
gain insight into the nucleosynthesis and element enrichment of the early
interstellar medium. For the majority of EMP stars the elements ratios follow
a normal pattern, but 10-15 \% of the stars are enhanced in heavy
neutron-capture (\emph{r}- or \emph{s}-process) elements by large factors, and about 20 \%
are strongly enriched in carbon \cite{2005}. 
The enhancement of some elements could be the result of highly non-spherical
supernovae and inefficient mixing in the early interstellar
medium. Alternatively, they could be due to mass transfer from a former AGB or
supernova companion binary, that has now evolved to a white dwarf or a neutron
star. If the latter explanation is true, we can detect these stars as
long-period binaries.

\section{Sample and Method}
The sample contains 40 stars, 17 of which are enhanced in \emph{r}-process elements,
eight with moderate enhancement $(0.3 < [$r/Fe$] < 1.0)$ and nine with the large
overabundances $([$r/Fe$] > 1.0)$. The remaining 23 stars are carbon enhanced
stars (CEMP), of these 12 show signatures of \emph{s}-process elements in their
spectra (CEMP-\emph{s}) while 8 have no such signatures (CEMP-\emph{no}). The last three
stars show both \emph{r}- and \emph{s}-process signatures (CEMP-\emph{r/s}). The sample includes
CS22892-052 which is enhanced in both carbon and \emph{r}-process elements \cite{Sneden2003}.

Medium-resolution low S/N spectra of the stars have been obtained over a
period of five years with the FIES spectrograph at the Nordic Optical
Telescope at La Palma. Cross-correlation have been performed to monitor the
radial velocity variations of the stars over the period with a precision of
100-300m/s. 
\newpage 
\section{Results}
\subsection{r-process enhanced stars}
\begin{center}
\begin{figure}[h]\centering
\includegraphics[scale=0.3]{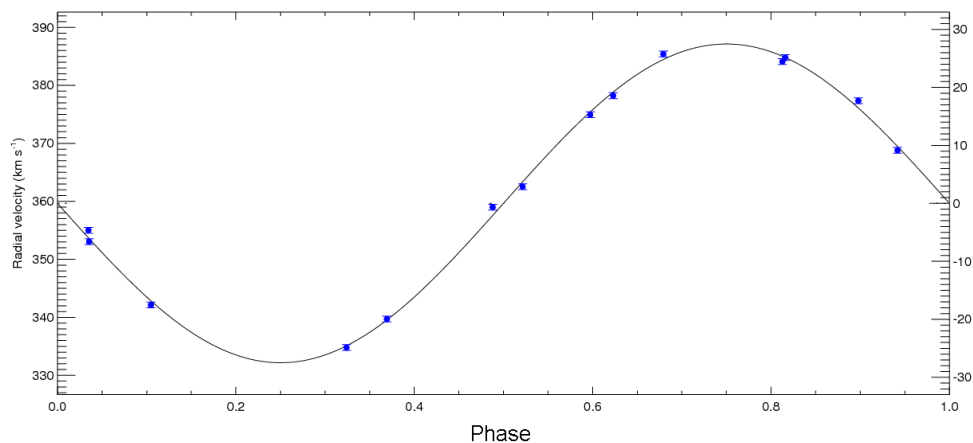}
\caption{Observed orbit of the \emph{r}-process enhanced star HE 1044-2509, with a period of 37 days.}
\label{r-star}
\end{figure}
\end{center}

Among the 17 \emph{r}-process enhanced stars three where found to be binaries \cite{Hansen2011},
the remaining 14 show no significant variation in their radial velocity over
the five year period. Figure \ref{r-star} shows the orbit of one of the
observed binaries, HE 1044-2509 with a period of 37 days. 
In all three binary systems found the stars are normal giant branch binaries
and there is no sign that the secondary has passed through the AGB stage of
evolution or exploded as a supernova, which could have poluted the surface of
the primary \cite{Mermilliod1996}.

\subsection{Carbon enhanced stars}
\begin{center}
\begin{figure}\centering
\includegraphics[scale=0.4]{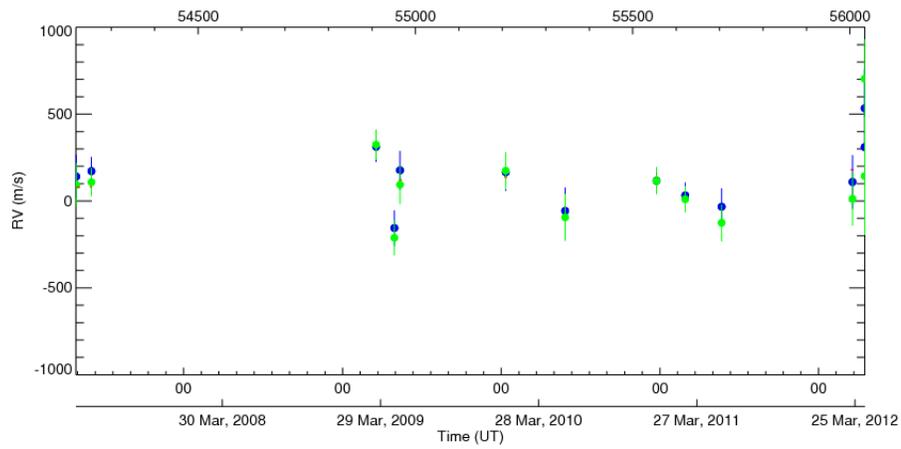}
\caption{Radial velocity measurements of the CEMP-\emph{no} star CS 22877-001
  over a period of 5 years, RMS=179.25m/s}
\label{CEMPno}
\end{figure}
\end{center}

\begin{center}
\begin{figure}\centering
\includegraphics[scale=0.4]{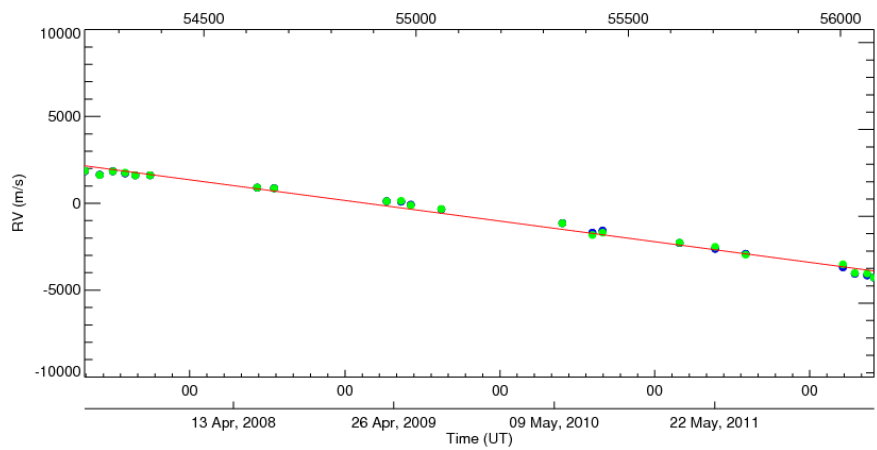}
\caption{Radial velocity measurements of the CEMP-\emph{r/s} star LP 625-44
  over a period of 5 years. RMS=2177.00 m/s}
\label{CEMPrs}
\end{figure}
\end{center}

\begin{center}
\begin{figure}\centering
\includegraphics[scale=0.3]{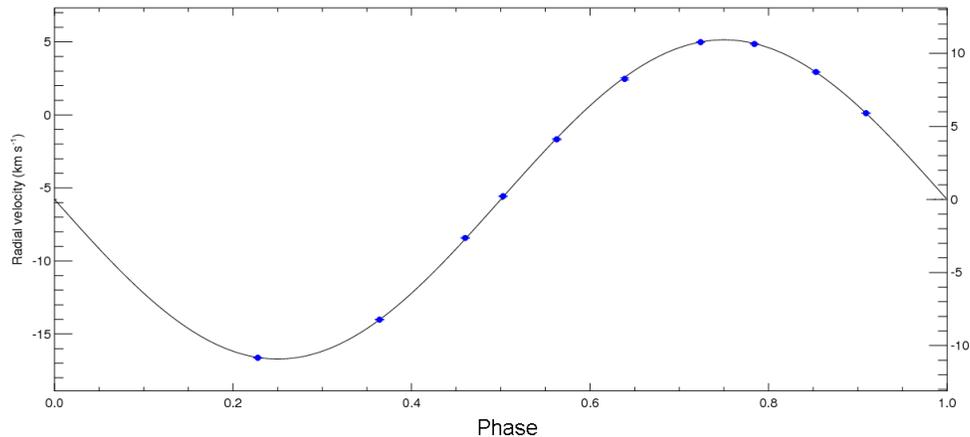}
\caption{Orbit of CEMP-\emph{s} star HE 0507-1430 with a period of 446 days.}
\label{CEMPs}
\end{figure}
\end{center}

The preliminary results for the carbon enhanced sample show that two out of
eight CEMP-\emph{no}, nine out of twelve CEMP-\emph{s} and one of the three CEMP-\emph{r/s} stars are
in binary systems. These results are shown in figures \ref{CEMPno}-\ref{CEMPs}. 
Figure \ref{CEMPno} shows the result of the radial velocity monitoring
of a CEMP-\emph{no} star. The plot clearly shows no significant variation in
radial velocities of the star over the five years of measurements. Figure \ref{CEMPrs} shows
the measured radial velocities for a CEMP-\emph{r/s} star, here there is a clear variation over the
five year period, indicating a binary with a very long period. Figure \ref{CEMPs} shows
an orbit plot of one of the binary CEMP-\emph{s} star systems. 
The majority of the CEMP-\emph{s} stars are found in binary systems, but not all of these are long period
binaries. Hence in some orbits there is no room for the secondary to have
passed through the AGB phase, where the mass transfer would take place.

\newpage
\newpage

\section{Conclusions}
The CEMP-\emph{no} and \emph{r}-process enhanced subsamples are found to have a normal
frequency of binary stars, this strongly suggests that the enhancement in
carbon and \emph{r}-process elements seen in these stars are not connected with the
binary status of the stars. The carbon and \emph{r}-process elements are extrinsic to
the stars observed today and were likely injected into the interstellar medium
in a collimated manner which make these stars chemical indicators of their
formation site in the early Galaxy. The majority of the CEMP-\emph{s} stars are found
to be part of a binary system, hence the origin of the carbon and s-process
elements in these stars could be due to mass transfer from an AGB companion,
but not all of the binary systems found have long periods with room to
accommodate an AGB star.


\begin{thebibliography}{99}
\bibitem{2005}T. C.~Beers \& N.~Christlieb, \emph{The Discovery and Analysis
  of Very Metal-Poor Stars in the Galaxy}, Ann. rev. A\&A {\bf 2005} (43) 531-580.
\bibitem{Hill2002} Hill et al. \emph{First stars. I. The extreme r-element
  rich, iron-poor halo giant CS 31082-001. Implications for the r-process
  site(s) and radioactive cosmochronology}, A\&A {\bf 2002} (387) 560-579. 
\bibitem{Francois2007} Fran{\c c}ois et al. \emph{First
  stars. VIII. Enrichment of the neutron-capture elements in the early
  Galaxy}, A\&A {\bf 2007} (476) 935-950.
\bibitem{Mermilliod1996} Mermilliod, J. C., \emph{Statistical Properties of
  Detected Binary Stars}, ASP Conf. Ser. {\bf 1996} (90) 95. 
\bibitem{Hansen2011} Hansen et al. \emph{The Binary Frequency of
  r-Process-element-enhanced Metal-poor Stars and Its Implications: Chemical
  Tagging in the Primitive Halo of the Milky Way}, ApJL {\bf 2011} (743) L1.
\bibitem{Sneden2003} Sneden et al. \emph{The Extremely Metal-poor, Neutron
  Capture-rich Star CS 22892-052: A Comprehensive Abundance Analysis}, ApJ {\bf 2003} (591) 936-953.
\end{thebibliography}
\end{document}